\documentclass[preprint,prb]{revtex4}

\usepackage{graphicx}
\usepackage{dcolumn}
\usepackage{float}
\usepackage{amsmath}

\newcommand{\comment}[1]{}

\begin{document}

\title{\boldmath Revisiting optical properties of MgB$_{2}$ with a high-quality sample prepared by a HPCVD method. \unboldmath}


\author{Yu-Seong Seo, Jae Hak Lee, Won Nam Kang, \& Jungseek Hwang$^{*}$}

\affiliation{Department of Physics, Sungkyunkwan University, Suwon, Gyeonggi-do 16419, Republic of Korea}

\date{\today}

\begin{abstract}

{\bf We investigated a high-quality MgB$_{2}$ thin film with a thickness of $\sim$1000 nm on an Al$_{2}$O$_{3}$ substrate using optical spectroscopy. We measured the reflectance spectra of the film at various temperatures both below, and above, the superconducting transition temperature, $T_c$ $\simeq$ 40 K. An earlier study showed that when the sample surface is exposed to air the optical properties of the surface change immediately, however, the saturated change is negligibly small in the far-infrared region. The optical conductivity spectrum in the normal state shows two (narrow and broad) Drude modes, with the narrow Drude mode being dominant in the low frequency region below 1000 cm$^{-1}$. Our study, which uses a good-quality sample, provides more reliable data on the optical properties of MgB$_2$, in a similar spectral range. The optical data is analyzed further using an extended Drude model, and the electron-phonon spectral density function, $\alpha^2F(\omega)$, is extracted. The spectral density function $\alpha^2F(\omega)$ features two peaks: a small one near 114 cm$^{-1}$, and a strong peak around the 550 cm$^{-1}$ where the B-B bond stretching phonon exists. Our data in the superconducting state does not show the expected energy shift of the onset of scattering associated with the $\alpha^2F(\omega)$ peaks.}
\\ \\

\noindent *Correspondence to [email: jungseek@skku.edu].

\end{abstract}


\maketitle

\section{Introduction}

Superconductivity in magnesium diboride (MgB$_2$), which has a superconducting transition temperature of $T_c$ = 39 K, was discovered in 2001\cite{nagamastu:2001}. The MgB$_2$ compound has a {\it P}6/{\it mmm} crystal structure\cite{nagamastu:2001}, and a multi-band electronic structure comprising $\sigma$ and $\pi$ bands. Owing to these multi-bands, MgB$_2$ exhibits two distinct superconducting gaps whose sizes are: $\sim$6.8 meV for the large $\sigma$ band gap, $\Delta_{\sigma}$; and $\sim$1.8 meV for the small $\pi$ band gap, $\Delta_{\pi}$\cite{choi:2002a,szabo:2001}. The superconductivity of MgB$_2$ is known to be of the Bardeen-Cooper-Schrieffer (BCS) type, which involves a strong electron-phonon interaction. It has also been shown that the electrons in the quasi-2D $\sigma$ bands, and the B-B bond-stretching phonons at $\sim$600 cm$^{-1}$ (i.e., $\sim$74 meV), are involved in the electron-phonon interaction\cite{kortus:2001,choi:2002a}. These electron-phonon interactions have been extracted from tunneling results\cite{dolgov:2003}, and existing optical results\cite{tu:2001,hwang:2014}. Although there are some previously reported optical data results for MgB$_2$ available in the literature\cite{tu:2001,kuzmenko:2002}, the optical data presented in these reports was obtained using poor quality samples, or samples with poor residual resistivity. A magneto-optical study on a MgB$_2$ single crystal has also been reported\cite{perucchi:2002}, but as their reported reflectance level at 50 cm$^{-1}$ is significantly lower ($\sim$70\%) than the measured reflectance spectra of MgB$_2$ at similar frequencies, it is possible that their absolute reflectance spectra data may not provide a true representation, as the authors used a mosaic of MgB$_2$ crystals, and might not have aligned their sample perfectly. Other studies have been conducted on the anisotropic optical properties of MgB$_2$ single crystals by Guritanu {\it et al}\cite{guritanu:2006}, and Kakeshita {\it et al}\cite{kakeshita:2006}, but the spectral ranges used in their work were limited 0.1 - 3.7 eV, and 0.075 - 3.0 eV, respectively.

In this report, we present new optical data obtained from a high-quality MgB$_2$ thin film sample, over a wide spectral range that includes the far-infrared (FIR) regime (60 - 8000 cm$^{-1}$). Our sample shows one-tenth of the residual resistivity compared to previously studied samples in the FIR region by optical spectroscopy\cite{tu:2001,kuzmenko:2002}. We note that Kuzmenko {\it et al.}\cite{kuzmenko:2002} studied a polycrystalline sample and Tu {\it et al.}\cite{tu:2001} studied a thin film sample prepared by pulsed laser deposition (PLD)\cite{kang:2001,tu:2001} on an Al$_{2}$O$_{3}$ substrate. In contrast, our new sample was prepared using a hybrid physical chemical vapor deposition (HPCVD)\cite{zeng:2002} method on an Al$_{2}$O$_{3}$ substrate. The HPCVD technique yields high-quality, epitaxial, MgB$_2$ thick films with a $c$-axis-oriented columnar structure\cite{seong:2007}. We compared our optical data with the data reported by Tu {\it et al.}\cite{tu:2001}, which we believe to be the most reliable {\it ab}-plane optical data in the FIR region available. It is noteworthy that the surface of a single crystal MgB$_2$ is very sensitive to air; one study showed that when the sample surface is exposed to air the ellipsometric parameters are affected immediately, due to a formation of a contamination\cite{guritanu:2006}. The authors also claimed that the changes in the parameters by the surface contamination were much smaller than the difference between the optical spectra of two samples grown under slightly different conditions. We think that the surface contamination occurs in a few top layers near the surface and does not affect much on the bulk optical properties of the sample in far-infrared region (for further details refer to the Supplementary information I). Detailed sample preparation and reflectance measurement technique are provided in the Methods section. At all measuring temperatures, our reflectance spectra yield higher values than those reported by Tu {\it et al.} in the FIR region. Interestingly, four sharp peaks below 750 cm$^{-1}$, which were observed and assigned as phonon modes of MgB$_2$ by Tu {\it et al.}, do not feature in our reflectance results. Our report concludes with a further analysis of the measured reflectance spectra and comparison of other optical quantities with those reported by Tu {\it et al.}.

\section{Results and discussion}

\begin{figure}[!htbp]
  \vspace*{-0.3 cm}%
  \centerline{\includegraphics[width=4.5 in]{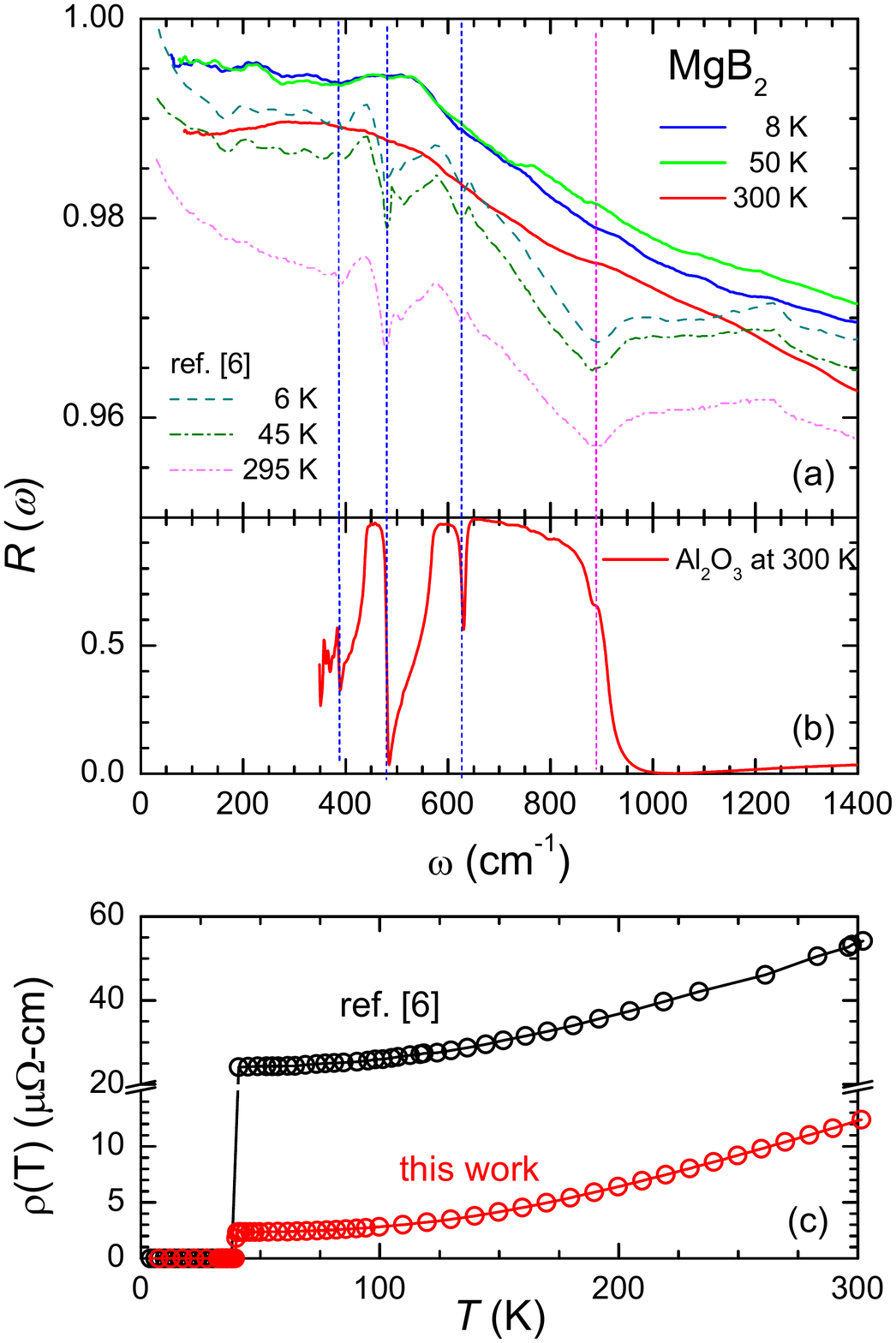}}%
  \vspace*{-0.7 cm}%
\caption{(a) Measured reflectivity spectra of our MgB$_2$ sample, along with previous MgB$_2$ data obtained by Tu {\it et al.}\cite{tu:2001}. (b) Measured reflectivity spectrum of the Al$_2$O$_3$ substrate at 300 K. (c) Measured resistivity of our MgB$_2$ sample, along with previous MgB$_2$ data by Tu {\it et al.}\cite{tu:2001}.}
\label{fig1}
\end{figure}

In Fig. 1(a), we display the measured reflectivity spectra (solid lines) from our MgB$_2$ thin film/Al$_2$O$_3$ sample, alongside the reproduced reflectivity spectra (dash-dotted lines) of Tu {\it et al.}. It is evident from the figure that our reflectance spectra are higher than Tu {\it et al.}'s, at the three selected temperatures. Moreover, we do not observe any significant enhancement in reflectance near the large superconducting gap energy, i.e., $2\Delta_{\sigma} =$ 13.6 meV (or 110 cm$^{-1}$). Although an earlier study has confirmed the existence of the superconducting (SC) gaps\cite{ortolani:2008}, the enhancement of reflectivity below the gap is less than 1\%, which is one of the reasons why observing the SC gap in optical experiments is nontrivial. Furthermore, our reflectance spectra do not show any significant phonon features, or other features near the 900 and 1250 cm$^{-1}$, compared with Tu {\it et al.}'s data. For comparison, we show the measured room-temperature reflectivity spectrum of our Al$_2$O$_3$ substrate in Fig. 1(b). The well-known IR-active phonon modes (i.e., the $E_u$ and LO modes)\cite{barker:1963} of Al$_2$O$_3$ appear clearly in the reflectance spectrum. Comparing the three reflectance spectra (i.e., those from the pair of MgB$_2$ thin film samples, and the Al$_2$O$_3$ substrate), one may suspect that Tu {\it et al.}'s data was contaminated by the Al$_2$O$_3$ substrate. However, Tu {\it et al.} have checked that the optical spectra from their 450 nm thick MgB$_2$ sample were free of contributions from the substrate's phonon modes.

In Fig. 1(c), we display the resistivity data of both our own MgB$_2$ film, and Tu {\it et al.}'s sample. Our sample shows much smaller resistivity than Tu {\it et al.}'s; the resistivity of our sample is approximately one-tenth the magnitude of Tu {\it et al.}'s near 50 K, and one quarter of their value at 300 K. Moreover, the residual-resistivity ratio (RRR) of our sample is $\sim$4.8, while that of Tu {\it et al.}'s sample is $\sim$2.5. We note that our sample (1000 nm) is roughly twice as thick as the one studied by Tu {\it et al.}'s (450 nm). We estimate the skin depths of the two MgB$_2$ samples using the measured resistivity data at 500 cm$^{-1}$. The skin depths of Tu {\it et al.}'s sample are $\sim$61 nm, and $\sim$90 nm at 45 K, and 300 K, respectively, while those of our own sample are $\sim$19 nm, and $\sim$42 nm at 50 K, and 300 K, respectively. The determined skin depths inform us that both samples are sufficiently thick enough not to allow light from passing through them. Through our simple simulations, we find that the four phonon peaks that are visible in Tu {\it et al.}'s results are greatly suppressed in our data, due to our sample having a stronger charge carrier screening effect than Tu {\it et al.}'s sample (for further details refer to the Supplementary information II).

Hwang and Carbotte\cite{hwang:2014} have also analyzed Tu {\it et al.}'s data\cite{tu:2001} by removing the four phonon modes, and extracting the electron-phonon spectral density function, $\alpha^2F(\omega)$, using a generalized Allen's formula\cite{allen:1971,shulga:1991,sharapov:2005} and maximum entropy method\cite{schachinger:2006}. To correctly simulate Tu {\it et al.}'s data close to 1300 cm$^{-1}$, the authors needed to introduce a non-constant density of states due to the spectral feature near the 900 cm$^{-1}$ (refer to Fig. 1(a)), whose origin was not yet known. However, this non-constant density of states is not a feature predicted from theory\cite{kortus:2001}. Such a result suggests that Tu {\it et al.}'s data may feature some unknown extrinsic components.

\begin{figure}[!htbp]
  \vspace*{-0.4 cm}%
  \centerline{\includegraphics[width=4.5 in]{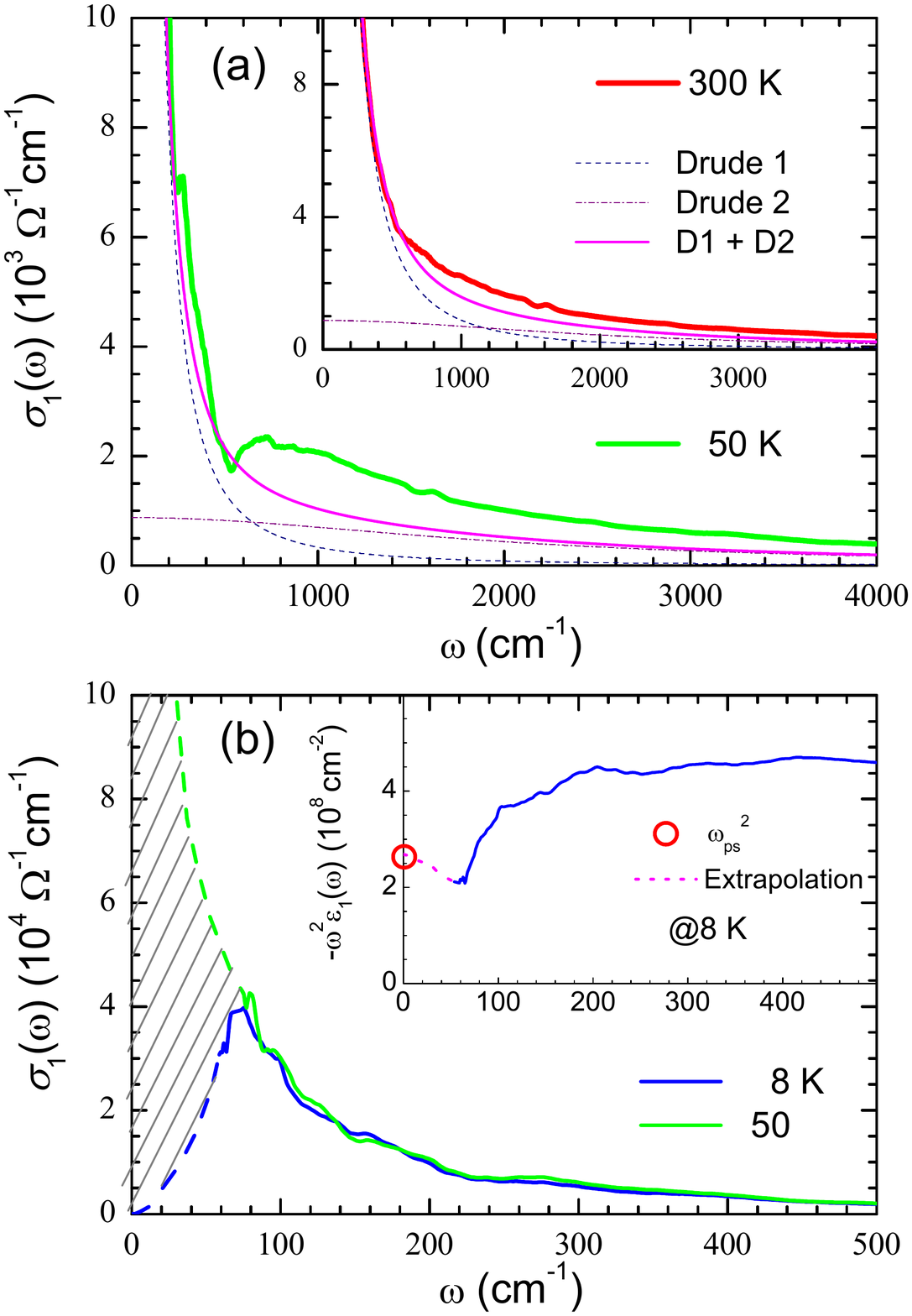}}%
  \vspace*{-0.9 cm}%
\caption{(a) The optical conductivity spectra at 50 K, with a theoretical fit using two Drude modes. In the inset we show the same spectra at 300 K. (b) The real part of the optical conductivity spectra in the superconducting state (8 K), and normal state (50 K). In the inset, we display the physical quantity $-\omega^2\epsilon_1(\omega)$ at 8 K, where $\epsilon_1(\omega)$ is the dielectric function as a function of frequency. The open red symbol is the superfluid plasma frequency squared, $\omega_{ps}^2$, obtained from the FGT sum rule\cite{glover:1956,ferrell:1958} (see main text).}
\label{fig2}
\end{figure}

In Fig. 2(a) we display the optical conductivity data for $\sigma_1(\omega)$ at 50 K, obtained from the measured reflectance using a Kramers-Kronig analysis and a fit with two (narrow and broad) Drude modes\cite{kuzmenko:2002,kakeshita:2006}; the narrow mode is known to be in the $\sigma$-band, while the broad mode is in the $\pi$-band\cite{kakeshita:2006}. We note that the fitting mismatch above the 570 cm$^{-1}$ is related to the incoherent part of the charge carriers which can appear due to electron-phonon interactions\cite{hwang:2008a}. From the fit, we estimate the Drude plasma frequencies, $\omega_{p,Di}$, and impurity scattering rates, $1/\tau_{imp,i}$ (where $i =$ 1 or 2), as $\omega_{p,D1} =$ 20,500 cm$^{-1}$ and $1/\tau_{imp,1} =$ 48 cm$^{-1}$ for the narrow $\sigma$-band, and $\omega_{p,D2} =$ 10,250 cm$^{-1}$ and $1/\tau_{imp,2} =$ 2000 cm$^{-1}$ for the broad $\pi$-band, respectively. The narrow Drude mode appears to be dominant in the low frequency region below 1000 cm$^{-1}$. In the inset to Fig. 2(a), we display the same data at a temperature of 300 K. The Drude plasma frequencies ($\omega_{p,Di}$) and impurity scattering rates ($1/\tau_{imp,i}$) at 300 K are 20,500 cm$^{-1}$ and 128 cm$^{-1}$ for the narrow $\sigma$-band, and 10,250 cm$^{-1}$ and 2000 cm$^{-1}$ for the broad $\pi$-band, respectively. The narrow Drude mode shows a 51\% larger plasma frequency, and a 36\% smaller scattering rate compared with the results of Tu {\it et al.} at 295 K. The difference in these results approximately explains the observed DC resistivity difference between the two samples at 300 K.

In Fig. 2(b) we display the real part of the optical conductivity at both $T=$ 8 K (i.e., for $T \ll T_c \approx 40$ K) and $T=$ 50 K (i.e., $T > T_c$). The missing spectral weight marked with hatched lines disappears from finite frequency region when the sample enters its superconducting phase. From this missing spectral weight, one can estimate the superfluid density, $n_s$, which is related to the superconducting plasma frequency, $\omega_{ps}$, by the relation: $4 \pi n_s e^2/m = \omega_{ps}^2$, where $e$ is the unit charge and $m$ is the electron mass. This missing spectral weight can be described by the well-known Ferrell-Glover-Tinkham (FGT) sum rule\cite{glover:1956,ferrell:1958}, $\omega^2_{ps} = 8\int_{0^+}^{\omega_c} [\sigma_{1,N}(\omega)-\sigma_{1,SC}(\omega)]d\omega$, where $\sigma_{1,N}(\omega)$ and $\sigma_{1,NC}(\omega)$ are the optical conductivities of the normal and superconducting states, respectively. From this missing spectral weight, we obtain the superconducting plasma frequency $\omega_{ps}$ = 16,400 cm$^{-1}$, and the corresponding London penetration depth of 97 nm. We note that since the missing spectral weight occurs mostly in the extrapolated region, the estimated London penetration depth can be considered as a lower limit. In general, if the cutoff frequency increases the estimated London penetration depth can decrease, since more missing spectral weight can contribute. To determine how the London penetration depth depends on the low-cutoff-frequency, we performed the same analysis using a higher low-cutoff-frequency of 80 cm$^{-1}$, obtaining a London penetration depth of 92 nm. Thus our estimated London penetration depth does not show a large dependence on the low-cutoff-frequency. We note that, in general, the London penetration depth also depends on the measured reflectance spectrum level. The corresponding values of the superconducting plasma frequency and London penetration depth determined from the Tu {\it et al.} sample\cite{tu:2001} are 7300 cm$^{-1}$ and 218 nm, respectively. The superconducting plasma frequency can also be estimated using the dielectric function, $\epsilon_1(\omega)$, since $\omega_{ps} \equiv \sqrt{\lim_{\omega \rightarrow 0} [-\omega^2\epsilon_1(\omega)]}$. The quantity $-\omega^2\epsilon_1(\omega)$ is displayed in the inset of Fig. 2(b) as a function of frequency. The open circle along the vertical axis corresponds to the superconducting plasma frequency obtained from the FGT sum rule. The SC plasma frequencies obtained from our two methods of analysis are in good agreement with one another.

We now analyze our data further using an extended Drude model formalism\cite{puchkov:1996,hwang:2004} and a generalized Allen's formulas\cite{allen:1971,shulga:1991} to study the electron-phonon interaction. The extended Drude model formalism allows one to describe correlated electrons effects, where information about the correlation appears in the so-called optical self-energy, $\tilde{\Sigma}^{op}(\omega)\equiv \Sigma^{op}_1(\omega)+i\Sigma^{op}_2(\omega)$\cite{hwang:2004}, which is closely related to the well-known quasiparticle self-energy\cite{carbotte:2005,hwang:2007b}. In the extended Drude formalism the optical self-energy can be written as, $\tilde{\Sigma}^{op}(\omega) =\omega/2 - i\omega_p^2/[8 \pi \tilde{\sigma}(\omega)]$, where $\omega_p$ is the plasma frequency, and $\tilde{\sigma}(\omega)= \sigma_1(\omega)+i\sigma_2(\omega)$ is the complex optical conductivity. The imaginary part of optical self-energy is related to the well-known optical scattering rate, $1/\tau^{op}(\omega)$, by $-2\Sigma^{op}_2(\omega) = 1/\tau^{op}(\omega)$, while the real part is related to the effective optical mass, $m^*_{op}(\omega)$ by $-2\Sigma^{op}_1(\omega) = [m^*(\omega)/m_b -1]\omega = \lambda^{op}(\omega)\omega$. Here, $m_b$ is the band mass, and $\lambda^{op}(\omega)$ is the optical coupling constant. Using generalized Allen's formulas, one can also extract the electron-phonon spectral density function, or the Eliashberg function, $\alpha^2F(\omega)$, from the optical-self energy, where $\alpha$ is the coupling constant between electron and the mediating phonons, and $F(\omega)$ is the phonon spectrum\cite{shulga:1991,sharapov:2005}. These generalized Allen's formulas are linear integral equations relating the optical self-energy to the electron-phonon (or electron-boson) spectral density function. Here, we used a generalized Allen's formula derived by Shulga {\it et al.}\cite{shulga:1991} in order to extract $\alpha^2F(\omega)$ from the measured optical scattering rate at 50 K. Furthermore, Shulga {\it et al.}'s formula can be applied to analyze optical data at a finite temperature with a constant density of states, and can be written as $1/\tau^{op}(\omega,T) = (\pi/\omega)\int^{\infty}_{0} \alpha^2F(\Omega)\{2\omega \coth(\Omega/2T) - (\omega+\Omega)\coth[(\omega+\Omega)/2T] + (\omega-\Omega)\coth[(\omega-\Omega)/2T]\}d\Omega + 1/\tau_{imp}$, where $T$ is the temperature and $1/\tau_{imp}$ is the impurity scattering rate. The electron-phonon spectral density function, $\alpha^2F(\omega)$, is determined by solving Shulga {\it et al.}'s equation numerically.

\begin{figure}[!htbp]
  \vspace*{-0.4 cm}%
  \centerline{\includegraphics[width=4.5 in]{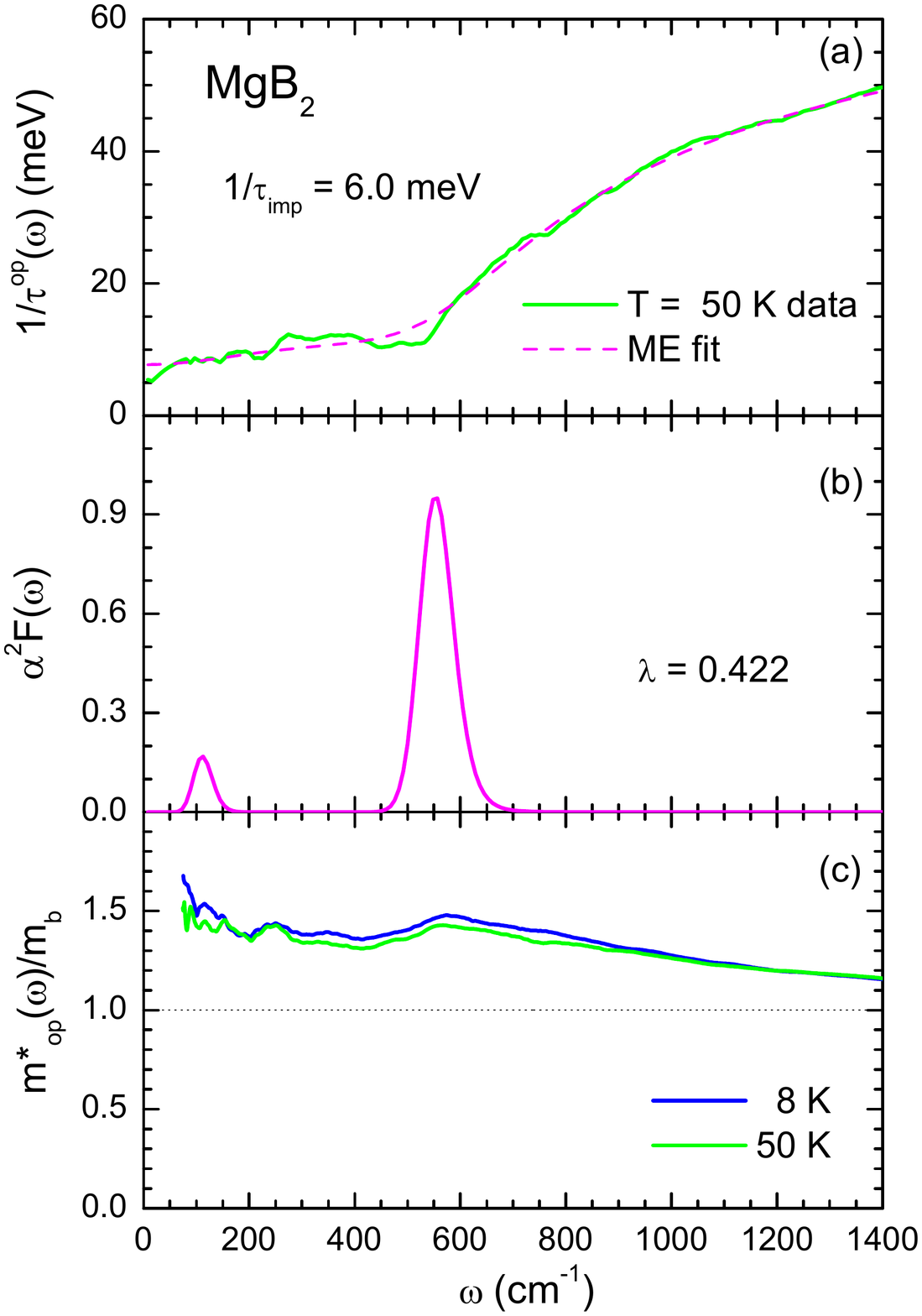}}%
  \vspace*{-0.9 cm}%
\caption{(a) The optical scattering rate of MgB$_2$ at 50 K and its maximum entropy (ME) fit, obtained from Shulga {\it et al.}'s formula\cite{shulga:1991} with the impurity scattering, $1/\tau_{imp}$ = 6.0 meV. (b) Extracted electron-phonon spectral density function, $\alpha^2F(\omega)$, from the ME fit. (c) The optical effective mass to band mass ration, $m^*(\omega)/m_b$, at temperatures $T =$ 8 K and $T =$ 50 K.}
\label{fig3}
\end{figure}

In Fig. 3(a) we display the optical scattering rate of our MgB$_2$ thin film sample at 50 K, along with a fit obtained using a maximum entropy (ME) method\cite{schachinger:2006} within Shulga {\it et al.}'s formalism\cite{shulga:1991}. Here, in order to obtain the optical scattering rate in the extended Drude model formalism, we took the plasma frequency to be 25,590 cm$^{-1}$, which includes the total spectral weight up to 6000 cm$^{-1}$, an impurity scattering rate $1/\tau_{imp}$ = 6.0 meV (or 48 cm$^{-1}$), and the background dielectric constant of $\epsilon_{H}$ = 1.6. We note that the ME fit is quite good up to 1400 cm$^{-1}$ and probably up to even higher frequencies, however the fit cannot capture the sharp kink near 550 cm$^{-1}$ completely; the experimental kink is sharper than the simulated one. The resulting electron-phonon spectral density function, $\alpha^2F(\omega)$, extracted using this analysis is displayed in Fig. 3(b), where two peaks are visible. The first peak is a weaker one and is located at the $\sim$114 cm$^{-1}$, while the second strong peak occurs at $\sim$550 cm$^{-1}$. We note that the weak peak at $\sim$114 cm$^{-1}$ is needed in order to capture an overall crease from 50 to 500 cm$^{-1}$. In our fit, we have ignored small sharp features in the scattering rate, like the feature around the 250 cm$^{-1}$. We also note that the extracted $\alpha^2F(\omega)$ includes contributions from both the $\sigma$- and $\pi$-bands. Our spectral density function $\alpha^2F(\omega)$ looks quite similar to that obtained in the previous study of Hwang and Carbotte\cite{hwang:2014}, however, there the authors needed to introduce a non-constant density of states to get a reasonable fit to the data of Tu {\it et al.}\cite{tu:2001} up to 1300 cm$^{-1}$. As we have already mentioned, this was because Tu {\it et al.}'s data contained additional extrinsic features near the 900 cm$^{-1}$ and 1250 cm$^{-1}$ (refer to Fig. 1(a)). The value of the coupling constant, $\lambda \equiv 2\int_0^{\omega_c}\alpha^2F(\Omega)/\Omega \: d\Omega \cong \lambda^{op}(0)$, obtained from the extracted $\alpha^2F(\omega)$ was determined to be  $\lambda =$ 0.422, which is similar to the result of an earlier study\cite{choi:2002}. In Fig. 3(c) we display the effective optical mass, $m^*_{op}(\omega)/m_b$, at the temperatures $T$ = 8 K, and $T =$ 50 K. The effective mass is consistent with the coupling constant, i.e., $m^*_{op}(0)/m_b \simeq 1 + \lambda$. We can also estimate the superconducting transition temperature from the extracted $\alpha^2F(\omega)$ using a generalized McMillan formula\cite{mcmillan:1968,carbotte:1990}, $T_c^{max} \simeq 1.13 \:\:\omega_{ln}\: exp[-(1+\lambda)/\lambda]$, where $T_c^{max}$ is the maximum SC transition temperature, and $\omega_{ln}$ is the logarithmically averaged frequency defined as $(2/\lambda) \int^{\omega_c}_0 \: \ln{\Omega}\:\: [\alpha^2F(\Omega)/\Omega] \:d\Omega $. The values estimated from this analysis are $\omega_{ln}$ = 40.7 meV, and a maximum transition temperature of $T_c^{max}$ = 18.4 K, which is significantly smaller than the measured $T_c \simeq$ 40 K. Thus, the extracted $\alpha^2F(\omega)$ does not seem to be strong enough to explain the SC transition temperature. However, if we use a higher plasma frequency in the extended Drude model analysis, we can obtain a larger value for $\lambda$ and increase the $T_c^{max}$. Indeed, if we increase the plasma frequency $\omega_p$ = 25,590 cm$^{-1}$ by 30\%, we will have a higher coupling constant value $\lambda$ = 0.71, and a maximum superconducting transition temperature $T_c^{max}$ = 48.0 K.

In order to simulate the measured optical scattering rate at 8 K (i.e., for $T \ll T_c \simeq$ 40 K) we used the extracted spectral density function $\alpha^2F(\omega)$ at 50 K and Allen's formula\cite{allen:1971} for the superconducting state. Allen's formula can be written in the form $1/\tau^{op}(\omega) = (2\pi/\omega)\int_0^{\omega-2\Delta}\alpha^2F(\Omega)(\omega-\Omega)E[\sqrt{1-4\Delta^2/(\omega-\Omega)^2}]d\Omega
+1/\tau_{imp}E[\sqrt{1-4\Delta^2/\omega^2}]$, where $\Delta$ is the SC gap, and $E(x)$ is the complete elliptic integral of the second kind. We note that contributions from both SC gaps, $\Delta_{\sigma}$ and $\Delta_{\pi}$, need to be included. As in the published literature\cite{dolgov:2003}, we assumed that the contributions from the two parallel SC channels of the $\sigma$- and $\pi$-bands comprised 0.33 and 0.67 of the total plasma frequency squared, respectively, and that the impurity scattering rate $1/\tau_{imp}$ = 6.0 meV (or 48 cm$^{-1}$) was the same as for the 50 K case. This simulation process is similar to an approach that has been applied to analyze the multiband FeAs superconductors\cite{hwang:2016} (see also the Supplementary information III). In Fig. 4(a) we display the measured optical scattering rate at 8 K (solid blue line), along with the simulated one (dashed orange line). The two SC gaps at 3.8 mev ($2\Delta_{\pi}$) and 13.6 meV ($2\Delta_{\sigma}$) are marked with magenta arrows. We have also marked the position of the major peak in $\alpha^2F(\omega)$ with a dark blue arrow, $\Omega_R \simeq$ 69 meV (or 553 cm$^{-1}$), and the two expected characteristic energy scales,  $\Omega_R + 2\Delta_{\pi}$ and $\Omega_R + 2\Delta_{\sigma}$ with red arrows. It can also be observed that the characteristic energy scales clearly in the derivative of the simulated optical scattering rate, shown by the grey line at the bottom of Fig. 4(a), where the vertical axis is measured in arbitrary units.

\begin{figure}[!htbp]
  \vspace*{-0.3 cm}%
  \centerline{\includegraphics[width=4.5 in]{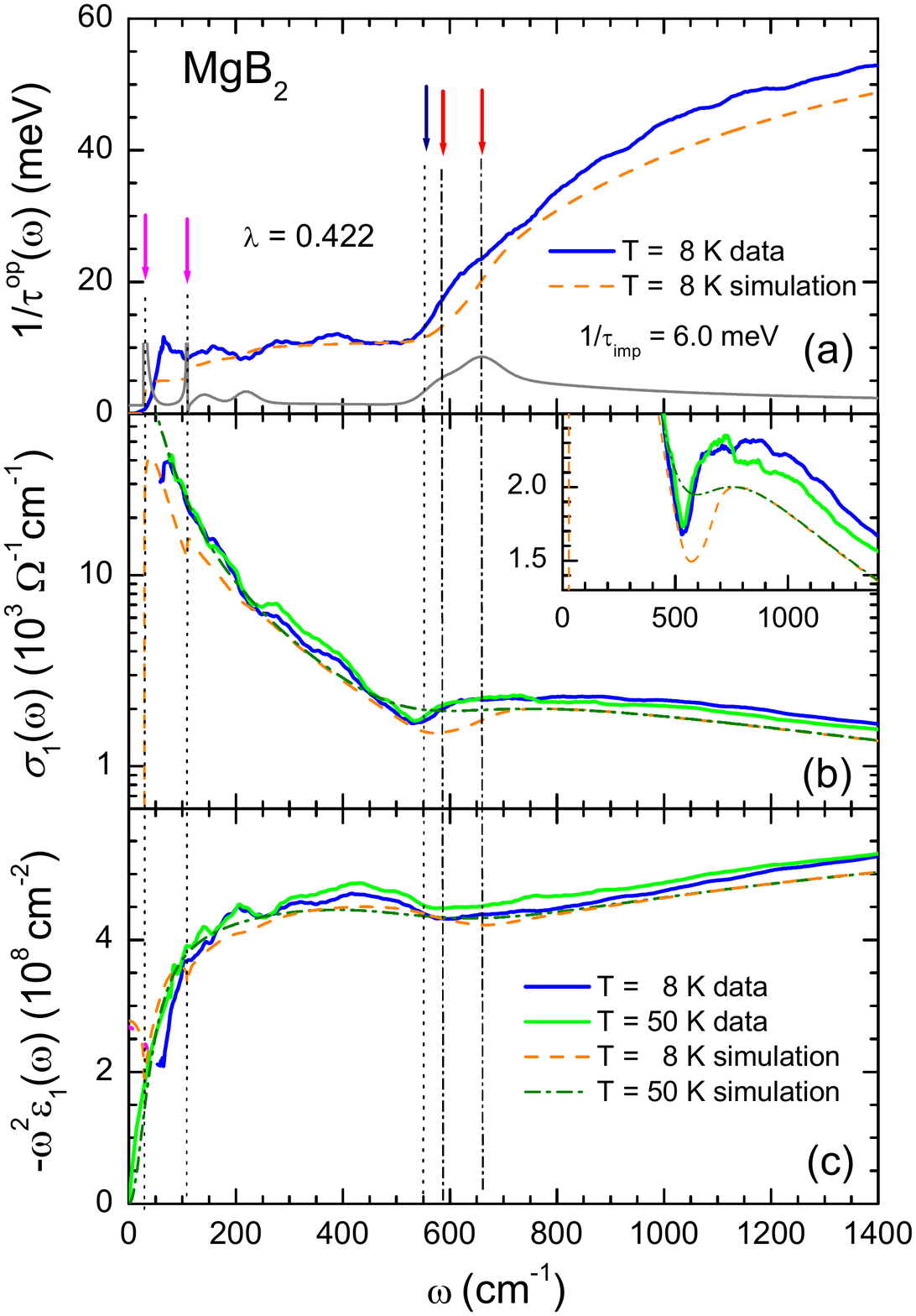}}%
  \vspace*{-0.8 cm}%
\caption{(a) The measured optical scattering rate, $1/\tau^{op}(\omega)$ at 8 K (superconducting state), and the optical scattering rate calculated using Allen's formula\cite{allen:1971} together with $\alpha^2F(\omega)$ extracted from the $1/\tau^{op}(\omega)$ data at 50 K through a maximum entropy (ME) fitting. The grey line is the derivative of the simulated optical scattering rate at 8 K. (b) The optical conductivity data at 8 K and 50 K and simulated values obtained using the extracted $\alpha^2F(\omega)$ through a reverse process\cite{hwang:2015a}. In the inset, we display a magnified view in order to show the dips of the measured and simulated data near 550 cm$^{-1}$. (c) The quantity $-\omega^2\epsilon_1(\omega)$ at 8 K and 50 K, and corresponding simulated values. The two magenta arrows mark the positions of the superconducting gaps: $2\Delta_{\pi}$ and  $2\Delta_{\sigma}$. The dark blue arrow marks the position, $\Omega_R$, of the strong (or major) peak of $\alpha^2F(\omega)$. The two red arrows mark the positions of the characteristic energy scales, $\Omega_R+2\Delta_{\pi}$ and $\Omega_R+2\Delta_{\sigma}$.}
\label{fig4}
\end{figure}

To further highlight the characteristic features of the optical data, we display both the measured and simulated values for the optical conductivity, $\sigma_1(\omega)$ and the quantity $-\omega^2\epsilon_1(\omega)$ at $T$ = 8 K and $T$ = 50 K in Figs. 4(b) and 4(c), respectively. We can clearly see a dip near 550 cm$^{-1}$ in both the measured and simulated conductivity data, which is a consequence of the major peak of $\alpha^2F(\omega)$; this dip also appears in the dielectric function. The dip in the optical conductivity divides the spectral weight into two parts: coherent and incoherent components\cite{hwang:2008a}. However, we note that since the optical conductivity and the dielectric function are independent optical quantities (though related through a Kramers-Kronig relation), the dip positions in the two different quantities are, in principle, not the same. Here, we also note that the measured spectral dip positions for both the 8 K and 50 K data are identical, with the shapes of the dips being almost identical too. In general, the dip at a SC state is sharper and shifted by twice of the SC gap, compared to that of a normal state\cite{hwang:2015a}. We further note that the reflectance spectra of Tu {\it et al.}\cite{tu:2001} do not appear to show such a shift and sharpening near the 550 cm$^{-1}$ either (refer to Fig. 1(a)). In the inset of Fig. 4(b) we show a magnified view of the dips in the measured and simulated curves for 8 K (dashed orange curve) and 50 K (dot-dashed dark green curve). Since the position of the major peak is very large compared with the two SC gaps (i.e., $\sim$10 times of the lager gap, $\Delta_{\sigma}$), and the two SC gaps ($\Delta_{\sigma}$ (6.8 meV) 33\% and $\Delta_{\pi}$ (1.8 meV) 67\%) contribute to the dip, the pair of dips are quite broad in appearance and have similar energy scales (see Supplementary information III). Although, we were able to explain some characteristic features (which might be closely related to the Cooper pairing) in the measured optical spectra of MgB$_2$ with Allen's model, we could not however, simulate similar sharpness of the dips in the measured data at 8 K and 50 K with our simple Allen's model.

\section{Conclusions}

We studied a high-quality MgB$_2$ thin film sample prepared by the HPCVD method, and provided experimental data for various optical properties of MgB$_2$. Furthermore, we analyzed our data using an extended Drude formalism in order to study the electron-phonon interaction, which is known to generate superconductivity in the material. We obtained the optical self-energy and subsequently extracted the electron-phonon spectral density function, $\alpha^2F(\omega)$ from the self-energy using a generalized Allen's formula derived by Shulga {\it et al.}\cite{shulga:1991}. The extracted $\alpha^2F(\omega)$ features two peaks, with the stronger peak yielding a sharp scattering onset near the 550 cm$^{-1}$ in the optical scattering rate. Interestingly, the onset is subject to a smaller shift than what is expected at the SC state. We attribute the small shift of the scattering onset energy to: (1) the high energy scale of the peak compared with the SC gaps, and (2) that both SC gaps jointly contribute to the onset. We believe that our new results supply more reliable optical property data than was previously available, and used a more pure MgB$_2$ sample. This experimental information on a two-gap superconducting system, MgB$_2$, may be relied on, and provide some insights to researchers in the field of multi-gap superconductivity.

\newpage

\section*{Methods}

\subsection*{Sample preparation and reflectance measurement technique.}
A high-quality MgB$_{2}$ thin film with a thickness $\sim$1000 nm was prepared on an Al$_{2}$O$_{3}$ substrate using a hybrid physical-chemical vapor deposition (HPCVD)\cite{zeng:2002,seong:2007,seong:2012}. The root-mean-square (RMS) surface roughnesses of our MgB$_2$ thin film samples are of the scale 3-4 nm, over a wide area of 3$\times$3 $\mu$m$^2$; we note that the RMS thickness of a well-optimized sample showed an RMS roughness as low as 1.5 nm\cite{seong:2012}. The superconducting transition temperature, $T_c$ of our sample is $\sim$40 K, and was obtained from the measured DC transport data. The thin film samples prepared by HPCVD show a high stability in air since remnant Mg elements do not exist on the sample surface. We note that there was a reported ellipsometry study on surface sensitivity of a single crystal MgB$_2$ to air; the authors showed that only top layers were contaminated\cite{guritanu:2006} (refer to Supplementary information I). For our optical study, we got a freshly prepared MgB$_2$ thin film sample, loaded it on a sample cone in a sample chamber, evacuated the sample chamber, and took the optical data. We used a commercial FT-IR spectrometer Vertex 80v, and a continuous flow liquid helium cryostat in order to obtain the reflectance spectra in the FIR and MIR range (60 - 8000 cm$^{-1}$), and at various selected temperatures from 8 to 300 K. In order to achieve accurate reflectance spectra we also applied an {\it in-situ} metallization method\cite{homes:1993}. In this method we used a coated 200 nm thick gold film on the sample as the reference reflectance. Furthermore, we corrected the measured reflectance with respect to the gold film by multiplying the measured value by the absolute reflectance of the gold film. We used an unpolarized beam for the reflectance measurement, with an incident angle on the sample of $\sim$10$^\circ$. We then performed a Kramers-Kronig analysis\cite{wooten} in order to obtain the optical constants, including the optical conductivity from the measured reflectance spectra which we extended to the high frequency region using previously reported reflectance data of MgB$_2$\cite{tu:2001}.

\bibliographystyle{apsrev4-1}
\bibliography{bib}

\noindent {\bf Author Contribution} J.H. and Y.S. wrote the main manuscript, Y.S. took the optical data and analyzed them. JH.L. and WN.K. prepared the thin film sample and took the dc resistivity data. All authors reviewed the manuscript.
\\ \\

\noindent {\bf Acknowledgements} J.H. acknowledges financial support from the National Research Foundation of Korea (NRFK Grant No. 2013R1A2A2A01067629). Y.S.S acknowledges financial support from the National Research Foundation of Korea (NRFK Grant No. 2016R1A6A3A11933016). This work was also supported by the Mid-career Researcher Program through an NRF grant funded by the Ministry of Education, Science \& Technology (MEST) (no. 2010-0029136). J.H. thanks Jules P. Carbotte for useful discussions to improve the paper.
\\ \\

\noindent {\bf Competing Interests} The authors declare that they have no competing financial interests.
\\ \\

\noindent {\bf Correspondence} Correspondence and requests for materials should be addressed to Jungseek Hwang~(email: jungseek@skku.edu).

\end{document}